\documentstyle[aps,prd]{revtex}
\def\bi{\bibitem}
\newcommand{\be}{\begin{equation}}
\newcommand{\ee}{\end{equation}}
\newcommand{\beq}{\begin{eqnarray}}
\newcommand{\eeq}{\end{eqnarray}}
\newcommand{\bear}{\begin{array}}
\newcommand{\ear}{\end{array}}

\begin{document}
\title{ The Complex Time WKB Approximation And Particle Production}    
\author{
S.Biswas$^{*a),b)}$, B.Modak$^{a)}$ and A.Shaw $^{**a)}$ \\ 
a) Dept. of Physics, University of Kalyani, West Bengal,\\
India, Pin.-741235 \\
b) IUCAA, Post bag 4, Ganeshkhind, Pune 411 007, India \\
$*$ email: sbiswas@klyuniv.ernet.in\\
$**$ email:amita@klyuniv.ernet.in}
\date{}
\maketitle
\begin{abstract}
The complex time WKB (CWKB) approximation has been an effective technique to 
understand particle production in curved as well as in flat spacetime. Earlier 
we obtained the standard results on particle production in time dependent gauge 
in various curved spacetime. In the present work we generalize the technique of 
CWKB to the equivalent problems in space dependent gauge. Using CWKB, we first 
obtain the gauge invariant result for particle production in Minkowski spacetime 
in strong electric field. We then carry out particle production in de-Sitter 
spacetime in space dependent gauge and obtain the same result that we obtained 
earlier in time dependent gauge. The results obtained for de-Sitter spacetime 
has a obvious extension to particle production in black hole spacetime. 
It is found that the origin of Planckian spectrum is due to repeated reflections 
between the turning points. As mentioned earlier, it is now explicitly shown 
that particle production is accompanied by rotation of currents.  
\end{abstract}
\newpage
\section{\bf{Introduction}}
The use of complex trajectories within the framework of Complex time WKB 
approximation (CWKB) has
been a recent trend \cite{bis:pram,bis1:pram,bis:cqg,guha:cqg,bis:ijmpa,sar:pram} 
to understand particle production in curved
spacetime as well as in Minkowski spacetime and can be extended to understand 
particle production from black holes. The method is 
also effective in quantum cosmology especially in evaluating the wavefunction of 
the universe \cite{bis:prd}, and to understand the preheating and reheating 
mechanism in 
inflationary cosmology \cite{sar:ph,shaw:pram}. Balian and Bloch \cite{bal:ann} 
formulated an approach to 
quantum mechanics starting from classical trajectories with complex coordinates. 
In the approach, within the framework of WKB approximation wave optics is 
generalized to complex trajectories to build up the wave. Such a method accounts 
for contributions of $\exp{(-c/\hbar)}$ to the usual WKB wave and is found to 
reproduce quantitatively all quantum mechanical effects even in cases where the 
scattering potential varies rapidly over a distance of wavelength or less. 
\par  
All complex trajectories are not allowed to contribute and one has to find a 
prescription for good (or, allowable) paths to be retained and how to combine 
them. It is question of topology in complex spaces, the topology of Stokes lines 
in one dimensional problems. In higher dimension one has to consider the 
topology of saddles, their position and height. The reader may consult Fedoryuk 
\cite{fed:asy} and Knoll and Schaeffer \cite{knoll:ann} for various aspects of 
Stokes lines, saddles 
and transition matrices within the framework of WKB approximation. The ref. 
\cite{knoll:ann} 
concentrates mainly on the heavy ion scattering, calculating reflection coefficient 
(identified as scattering amplitude) using the roles played by Stokes lines and 
saddles.
\par  
We generalized the technique of \cite{knoll:ann} for complex time considering 
Schroedinger 
like equation not in space but in time and find surprising and remarkable results 
when applied to various expanding spacetimes. It is well known fact that in 
Schroedinger like equation in time (i.e., the potential is also time dependent), 
the vacuum state at 
$t\rightarrow -\infty$  and at $t\rightarrow +\infty$ 
are not the same and is the root cause that affects particle production. 
\par
To understand particle production in such a situation, consider a potential 
$V(t)$ represented as a blob from which a pair is created; both particle and 
anti-particle of positive energy are moving forward in time away from the blob. 
Using Feynmann-Stuckleberg prescription, we identify negative energy 
particle propagating backward in time as being equivalent to positive energy 
anti-particle solution moving forward in time. Thus reflecting the direction of 
electron coming out of the blob, we interpret the pair production as reflection 
of positron off the potential $V(t)$.
\par  
The particle production in flat spacetime in an uniform electric field with 
$A_{\mu}=(0,Et,0,0)$ 
has been treated in \cite{bis:ijmpa} using CWKB and find the standard 
result. The more familiar potential
${A'}_{\mu}=(-Ex,0,0,0)$ 
is related to $A_\mu$ by a gauge transformation 
${A'}_\mu=A_\mu+\partial_\mu \Lambda$
with $\Lambda=-Ext$. The result of particle production should be independent of
the gauge i.e., should be the same for the choice $A_\mu$ or ${A'}_\mu$. In the 
space dependent gauge (i.e., for the choice ${A'}_\mu$), the vacuum of the field 
remains the same for all time and hence no particle production can take place. 
To recover the standard result in space dependent gauge, we use CWKB with a 
tunneling interpretation and obtain the gauge independent result. The method of 
complex paths, enunciated by Landau in \cite{lan:quant}, has been recently 
used in \cite{sri:iuc} to 
calculate the transmission and reflection coefficients for the equivalent 
quantum mechanical problem. Our result for reflection and transmission 
coefficients in CWKB differs from \cite{sri:iuc} but the expression for Bogolubov 
co-efficients and vacuum persistence probability coincide with that of gauge 
invariant method proposed by Schwinger. In (14), the Bogolubov coefficients 
$\alpha$ and $\beta$ are determined from a heuristic interpretation of unitarity 
relation
${\vert R \vert}^2+{\vert T\vert}^2=1$
in terms of Bogolubov coefficients that differs in space dependent and time 
dependent gauges. However in CWKB, we employ a different procedure 
\cite{bis:cqg,guha:cqg,bar:ictp} and the 
Bogolubov coefficients are obtained in a gauge invariant way. The advantage of
CWKB is that we do not require a knowledge of mode solutions and the analytic 
continuation 
$(x,t)\rightarrow -\infty$ to $(x,t)\rightarrow +\infty$ as has been done in other
works \cite{sri:iuc,bro:phy}. We also do not need a prior knowledge of 
the gauge invariant 
results. 
\par
Being encouraged by the success of CWKB in space dependent gauge we next consider
the de-Sitter spacetime and modify the previous work in the 
light of CWKB. We obtain the standard results usual to blackhole spacetime.
The details will be placed elsewhere. We mention the main results obtained 
only for de-Sitter spacetime.
\par
The basic problem in CWKB is to find the behaviour of a wave in a given region 
under the constraint of boundary conditions in other regions. During evolution, 
one encounters regions in which the particle (i.e., wavepacket) moves in a 
region of complex domain that is classically inaccessible and is characterized 
by turning points or saddles depending on the dimensions; but quantum mechanically 
there is a probability of transition in the classically unallowed region with a 
heuristic interpretation of tunneling. The heuristic interpretation gives a 
steady state of particle production. The similar situation exists in blackhole 
problems but here the turning points and saddles are replaced by horizons. The 
existence of horizon creates a separation in spacetime into portions very much 
akin to the behaviour of turning points such that straight trajectories become 
curved so as to approach or recede from the horizon exponentially slowly as seen 
by an external observer. A good review of black hole quantum physics is given in 
\cite{bro:phy}. Unfortunately, it requires a mastery in quantum physics as well as in 
classical relativity to understand the conceptual issues raised by quantum 
mechanics in the presence of horizon.
\par
The present paper is a critical review of particle production in curved 
spacetime and is a generalization of our previous works 
\cite{bis:pram,bis1:pram,bis:cqg,guha:cqg,bis:ijmpa} with an extension 
to blackhole problems. This is done especially to understand the blackhole 
evaporation and to settle some of the concepts introduced in \cite{bro:phy}, 
that we find 
different in CWKB. The CWKB has
the advantage to be applicable in the same way to spinor particle production 
and 
has been carried out in \cite{bis:cqg,guha:cqg,sar:pram}. The particle 
production in a time dependent
potential is understood as follows. A purely positive frequency wave with amplitude
$T$ in the infinite past $t\rightarrow -\infty$ evolves into a combination of 
positive and negative frequency waves in the infinite future 
$t \rightarrow \infty$ with negative frequency waves having an amplitude $R$ and 
positive frequency waves with amplitude unity. The process is viewed as a 
reflection in time as mentioned earlier. During discussion on spinor particle 
production \cite{bis:cqg,guha:cqg,bis:ijmpa} it was advocated that the pair 
production takes place due to 
rotation of currents from $-\vert J \vert$ to $+\vert J \vert$ as $t$ evolves 
from $-\infty$ to $+\infty$ and the mixing of positive frequency and negative 
frequency states occurs in the region between the turning points. The particle 
(wavepacket) moves then in complex $t$-plane. We verified this conclusion by 
carrying out numerical calculations in various expanding spacetimes 
\cite{sar:ijmpa,sar1:pram} with 
some interesting conclusions.
\par
Many results obtained in the present work are new. The organization of the paper 
is follows. In section 2 we discuss the basics of CWKB applicable both in the 
time dependent and space dependent gauge. In section 3 we discuss the CWKB 
particle production in Minkowski spacetime in an uniform electric field. In 
section 4. we clarify the heuristic interpretation of particle production 
through rotation of currents and from charge conservation and obtain the gauge 
invariant result even in CWKB. In section 5 we consider de Sitter spacetime in time dependent 
gauge and obtain the gauge invariant result. This example has an obvious
extension to blackhole spacetimes.
\smallskip
\section{\bf{Basics of CWKB}}
The details of complex time WKB approximation have been done in our previous 
works 
\cite{bis:pram,shaw:pram}. We mention here briefly for a follow up of the text of the paper. 
Let us consider a one dimensional Schroedinger equation 
\be
{\partial^2\over{\partial\eta^2}}+\omega^2(\eta)\psi=0.
\ee
The variable $\eta$ may correspond to space or time-like variable. In CWKB we
consider $\eta$ to be a complex variable and assume that $\omega(\eta)$ has complex 
turning points or $\omega(\eta)$ is complex between the two turning points (may be 
real). The turning points are given by
\be
\omega^2(\eta_{1,2})=0.
\ee
We consider the case for two turning points. The method can also be generalized 
for higher number of turning points. Defining
\be
S(\eta_f, \eta_i)= \int^{\eta_f}_{\eta_i} \omega(\eta) d\eta,
\ee
the solution of (1) in CWKB is written as 
\be
\psi (\eta)\rightarrow \exp{[iS(\eta, \eta_0)]} -iR\exp{[-iS(\eta, eta_0)]},
\qquad {\eta\rightarrow\infty}
\ee
Here $\eta_0$ and $\eta$ are real where $\eta_0$ is arbitrary and 
$\eta_0 > \eta$. We have neglected the WKB prefactor for simplicity. In (4) $R$
is given by
\be
R={{\exp{[2iS(\eta_1, \eta_0)]}}\over {1+\exp{[2iS(\eta_1, \eta_2)]}}}.
\ee
The derivation of (5) from global analysis in complex $\eta$-plane will be found 
in Knoll and Schaeffer \cite{knoll:ann}. We give here a transparent and 
intuitive 
derivation. Let a wave (to be identified as positron) starts from  large and 
positive $\eta_0$, and moves leftward to reach a point $\eta$ such that 
$\eta < \eta_0$. This is represented by the first term in (4). The second term 
is the reflected part contribution. A wave starts at $\eta=\eta_0$, moving 
leftward reaches the turning point $\eta_1$ and bounces back from $\eta_1$ 
rightward to reach $\eta_0$. It is represented as 
\be
(-i)\exp{[iS(\eta_1, \eta_0) -iS(\eta, \eta_1)]}.
\ee
Here $(-i)$ factor is introduced due to reflection. The contribution (6) is 
then multiplied by repeated reflection between the turning points $\eta_1$ and 
$\eta_2$ and is written as
\be
\sum_{\mu=0}^{\infty}{[-i\exp{\{iS(\eta_1, \eta_2)\}^{2\mu}}]} = {1\over 
{1+\exp{[2iS(\eta_1, \eta_2)]}}}.
\ee
The combined contributions (6) and (7) now comprise the second term in (4). 
We have used  
$iS(\eta, \eta_1)=iS(\eta, \eta_0)-iS(\eta_1, \eta_0)$ using relation (3). The 
boundary condition taken is, at $\eta\rightarrow -\infty$
\be
\psi(\eta)\rightarrow T e^{iS(\eta, \eta_0)},
\qquad {\eta\rightarrow -\infty} 
\ee
The interpretation of (4) and (8) now runs as follows. If we consider 
$\exp{[iS(\eta, \eta_0)]}$ as the anti-particle solution, (4) is interpreted as 
the reflection of a anti-particle from the turning point $\eta_1$ and the 
reflected part is interpreted as particle moving forward in time. Here $R$ and 
$T$ refer to reflection and transmission coefficient. The essence of (4) and (8) 
is that there is no particle at 
$\eta\rightarrow -\infty$, but at $\eta\rightarrow +\infty$, (8) evolves into 
(4) and is identified as a process of pair production where the pair production 
amplitude $\equiv$ reflection amplitude. It is interesting to note that the 
similarity of (4) with the ubiquitous moving mirror example \cite{bir:quan} and 
the origin of 
thermal spectrum. Considering $\eta$ as a time-like variable, we have been able 
to show using (4), (5) and (8) that in deSitter spacetime the particle creation 
occurs due to Hartle-Hawking vacuum decay, we also calculated the vacuum 
instability and found
$2Im L_{eff}={{4H^4}\over \pi^2}\exp{({-m\over {T_H}})}$
where $T_H={H\over {2\pi}}$ is the Hawking-deSitter temperature.                                                             
The approach given here can also be generalized to problems when $\eta$ is a
space like variable and the potential contains a space dependent part. We take a 
standard example of particle production in an uniform electric field in a  
space dependent gauge.
\smallskip
\section{\bf{CWKB Particle Production In An Uniform Electric Field}}
In presence of an e.m. field the minimally coupled scalar field $\phi$ 
propagating in flat spacetime satisfies the Klein-Gordon equation
\be
[(\partial_\mu+iqA_\mu)(\partial^\mu+iqA^\mu)+m^2]\Phi=0.
\ee
Let us take $A^\mu = (-E_0x,0,0,0)$. The electric field is $\vec{E}=E_0\hat{x}$.
We write
\be
\Phi= e^{-i\omega t}e^{ik_yy+ik_zz} \phi (x).
\ee
Substituting (10) in (9) we find
\be
{{\partial^2 \phi}\over {\partial x^2}} 
+[(\omega+qE_0x)^2 -M^2]\phi=0.
\ee
In (11) we have taken
\beq
{k_\bot}^2 &=& k_y^2+k_z^2, \nonumber \\
M^2 &=& k_\bot^2+m^2.
\eeq
The equation (11) is of the form (1) with the turning points given by
\be
(\omega+qE_0x)=\pm M.
\ee
It is better to have an convenient form of (11) and (13) making the following 
change of variables 
\be
\rho=\sqrt{qE_0}x+{\omega\over {q\sqrt{E_0}}}, \lambda={M^2\over {qE_0}},
\ee
in (11). We get
\be
{{\partial^2 \phi}\over {\partial x^2}}+(\rho^2-\lambda)\phi=0,
\ee
so that the turning points are at $\rho=\pm\sqrt{\lambda}$. Before
we apply CWKB to (15), let us try to understand the origin of complex paths 
in the formalism. Let us restrict to $(1+1)$ dimension and identify 
$x_c=-{\omega\over {qE_0}}$, and the acceleration $a={qE_0\over M}$ (now
$M=m$ since $k_\bot^2=0$). The classical trajectory is now given by
\be
(x-x_c)^2-(t-t_c)^2=a^{-2}
\ee
whereas the equation that determines the turning point reads
\be
(x-x_c+a^{-1}) (x-x_c-a^{-1})=0,
\ee
with $x_\pm =x_c\pm a^{-1}$ being the two turning points. Considers a value 
$x$ between the two turning points i.e.,
$x_c-a^{-1}<x<x_c+a^{-1}$. We see from (17) that
\be
(x-x_c)^2-a^{-2}<0.
\ee
Using (16), we now find
\be
(t-t_c)^2<0
\ee
i.e., $t$ becomes complex. This again implies from (16) that $(x-x_c)^2<0$ 
provided the electric field is large enough, a necessary condition of pair
production. Since the potential in (11) is space dependent we have to consider 
the motion in $x$ plane like a scattering problem in quantum mechanics. The 
conditions (4) and (8) are now replaced by
\be
\psi(x)\rightarrow e^{-iS(-x,x_0)}-iRe^{+iS(-x,x_0)} \qquad 
{x\rightarrow -\infty}.
\ee
As $x\rightarrow +\infty$, (20) evolves into
\be
\psi(x)\rightarrow T e^{-iS(x, x_0)},
\ee
where $R$ is now given by
\be
R={{e^{-2iS(x_1, x_0)}}\over {1+e^{-2iS(+x_1, +x_2)}}}.
\ee
In (21) and (22), $R$ and $T$ are respectively the reflection and 
transmission coefficients. We have now taken
\be
x_1=-\sqrt{\lambda}, x_2=+\sqrt{\lambda}.
\ee
In (21) and (22), $x_0$ is arbitrary. Its effect is to multiply (21) and (22)
by a factor $\exp{[i\delta (x_0)]}$ where $\delta (x_0)$ is real, so that 
$\vert R \vert^2$ and $\vert T\vert^2$ do not depend on $x_0$. So we neglect                                                                        
its contribution in evaluating $\vert R\vert$ and $\vert T \vert$. However, in 
present example we retain it to show our claim. Going from $x$ to $\rho$ variable
we now have
\be
S(\rho_f, \rho_i)=\int_{\rho_i}^{\rho_f}{(\rho^2-\lambda)^{1/2}\, d\rho}.
\ee
Remembering the fact that between the turning points, $(\rho^2-\lambda)^{1/2}$
is complex, we write
\be
S(\rho_f, \rho_i)=+i \int_{\rho_i}^{\rho_f}{(\lambda-\rho^2)^{1/2}\, d\rho}.
\ee
Taking $\rho_1 =-\sqrt{\lambda}$ and $\rho_2=+\sqrt{\lambda}$, we find
\beq
S(\rho_1, \rho_0)&=&-{i\lambda\pi\over 4} \nonumber \\
S(\rho_1, \rho_2)&=&{-i\lambda\pi\over 2}.
\eeq
Hence
\be
R= {{e^{i\,\delta (\rho_0)}}\,e^{{-\pi\,(\,k_\bot^2+m^2\,)}\over {2qE_0}}\over
{1 + e^{{-\pi\,(\,k_\bot^2+m^2\,)}\over {qE_0}}}}
\ee
where $\delta (\rho_0)$ is real. The expression (27) differs from the work
of Brout et. al \cite{bar:ictp} and \cite{sri:iuc}. It is therefore necessary 
to work out the
interpretation of (21), (22) and (27) in the light of CWKB and obtain the
standard Schwinger Gauge invariant result \cite{sch:pr}.
\section{\bf{Charge Conservation And Rotation Of Currents}}
Looking back to (21) and (22) we find for $\Phi$ given in (10) [suppressing
\, $\exp{(ik_yy+ik_zz)}$\, factor] as
\be
\matrix{\Phi_i  \cr\Phi_R \cr \Phi_T}  \left\}=
e^{-i\omega t}\right\{ \matrix{ e^{-iS(+x)},&\quad  x\rightarrow -\infty 
\cr Re^{+iS(+x)},&\quad x\rightarrow -\infty 
\cr Te^{-iS(+x)},&\quad x\rightarrow +\infty}
\ee
Writing $S(x)$ for large argument we get  
\be
\matrix{\Phi_i\cr\Phi_R\cr\Phi_T}\left\}=
e^{-i\omega t}\right\{\matrix{ 
e^{-iqE_0{{{(x+{\omega\over {qE_0}})^2}\over 2}}}
\vert x+{\omega\over {qE_0}}\vert^{ -{{im^2}\over {2qE_0}}-{1\over 2}}, 
&\quad {x\rightarrow -\infty}\cr
R e^{iqE_0{{{(x+{\omega\over {qE_0}})^2}\over 2}}}
\vert x+{\omega\over {qE_0}}\vert^{-{{im^2}\over {2qE_0}}-{1\over 2}}, 
&\quad{x\rightarrow -\infty}\cr
T e^{iqE_0{{{(x+{\omega\over {qE_0}})^2}\over 2}}}
(x+{\omega\over {qE_0}})^{-{{im^2}\over {2qE_0}}-{1\over 2}}, 
&\quad{x\rightarrow +\infty}}.
\ee
To obtain the prescription in $t$-space where $\phi_i, R, T$ live, we
differentiate the total phase in (29) with respect to $\omega$ and set
${{\partial S (total)}\over {\partial \omega}} =0$. For $\phi_i$ we get
\be
{{\partial S (total)}\over {\partial \omega}}=-(x+{\omega\over {qE_0}})-t
+{1\over {E\vert (x+{\omega\over {qE_0}})\vert}}=0.
\ee
From (30) we see that as $x\rightarrow -\infty ,\, t$ goes to $+\infty$. 
In a similar way, we obtain the $t$-behaviour of $\phi_{\sc{R}}$ and 
$\phi_{\sc{T}}$. 
Thus we find
\beq
\psi_i&\equiv&I (x\rightarrow -\infty, t\rightarrow +\infty) \nonumber \\
\psi_R&\equiv& R(x\rightarrow -\infty, t\rightarrow -\infty)\nonumber \\
\psi_T&\equiv& T(x\rightarrow +\infty, t\rightarrow +\infty).
\eeq
This behaviour is also obtained in \cite{bar:ictp} and is required to obtain the 
charge prescription in different regions.
Thus $\psi_i$ corresponds to motion to the left at late times where $R$
describes the motion to the right at early times and $T$ to the right but at
late times. Looking at (31) we note that at $t\rightarrow +\infty$ we have wave
moving forward and another wave moving backward in time and is identical in 
prescription discussing pair production through time dependent gauge. It 
establishes also the heuristic interpretation of particle-antiparticle rotation 
through tunneling mechanism. 
\par
Let us now look at the charge assignment of various branches. For the purpose
we construct the charge density using (29) and neglect
$\exp{(i\vec{k_\bot}.\vec{r_\bot})}$ term for convenience in interpretation. We get
\beq
J^t &=& \phi_\omega^* iD_t\phi_\omega
=\phi^*_\omega(i\stackrel{\longleftrightarrow}{\partial_t}+2qE_0x) \phi_\omega 
\nonumber \\
&=&2(\omega+qE_0x) \chi_{\sc{WKB}}^* \,\,\chi_{\sc{WKB}}
\eeq
where $\chi_{\sc{WKB}}$ is given in (29). At the turning point $\omega+qE_0x =-M,
J^t$ is negative. Thus both $\psi_i$ and $\psi_R$ that live at negative $x$
have negative charge. This is also exemplified by (32) when $x\rightarrow
-\infty$. Similarly at $\omega+qE_0x =+M$, we get $J^t$ positive  i.e.,
$\psi_{\sc{T}}$, that live at $x\rightarrow +\infty$ has positive charge. In
otherwords, the current $J^t$ undergoes a rotation from negative to positive
values as it evolves from $x\rightarrow -\infty$ to $x\rightarrow +\infty$.
This rotation occurs due to mixing of positive and negative frequency 
solutions within the turning points as envisioned earlier.
Now invoking the charge conservation
\be
\int{d^3x J^t\vert_{t=-\infty}}=\int{d^3 x J^t\vert_{t=+\infty}}
\ee
we get
\beq
-\vert R\vert^2&=&+\vert T\vert^2 -1\nonumber \\
\qquad \rm{or},\, \vert R\vert^2+\vert T\vert^2 =+1.
\eeq
In deriving (34) we have noted the fact $\psi_i$ and $\psi_{\sc{T}}$ live at
$t=+\infty$, whereas $\psi_{\sc{R}}$ lives at $t=-\infty$. Thus we see that the
unitarity relation itself is a statement of charge conservation.
Thus we observe from (31) that, though we worked with a space dependent gauge,
we arrive at a situation i.e., a pair at $t\rightarrow +\infty$ and no
particle at $t\rightarrow -\infty$ and which is identified as pair production
via (4) and (8). Thus we arrive at the same interpretation in both the gauges.
To put (34) in terms Bogolubov co-efficients $\alpha$ and $\beta$, 
Brout et. al \cite{bar:ictp}
identify
\be
\vert \alpha\vert^2={1\over {\vert R\vert^2}}, \qquad \vert\beta\vert^2=
{{\vert T\vert^2}\over {\vert R\vert^2}}
\ee
so that $\vert \alpha\vert^2-\vert\beta\vert^2=1$. However our result for
$\vert R\vert^2$ and $\vert T\vert^2$ differs from \cite{bar:ictp} and 
Padmanabhan \cite{sri:iuc}
and naturally the Bogolubov co-efficients will be different and would not get
the Schwinger result if we follow (35). In our view (35) is not a correct 
choice or is an
approximate relation. We know that Bogolubov co-efficients find an
interpretation only through
\be
(\rm{number\; \;of\; pairs\; created\; in\; a\; mode\;} k\,)\equiv 
\vert\beta_k\vert^2
\ee
This is the basic definition and interpretation of $\vert \beta_k\vert^2$.
To establish the correctness of CWKB, we now turn to the calculation of vacuum
persistence probability to investigate the correctness of our approach or
feasibility of the expression (35). It needs a calculation of $\alpha_k$ and
$\beta_k$.
\par
In CWKB, we calculate $\beta_\omega$ and $\alpha_\omega$ through a different
approach. The reflection co-efficient $\vert R\vert^2$ is identified as the
probability for the creation of one pair with wave number $\vec{k}(\equiv klm)$.
We write it as [3,19]
\be
\vert R\vert^2=\vert A_0\vert^2\omega_k
\ee
where $A_o$ is the amplitude for no particle production and $\omega_k$ is the
relative probability of creating a pair in a given mode. The probability of
producing $n$ pairs with wavenumber $\vec{k}$ is
\be
P_n(k)=\vert A_0\vert^2\omega_k^n.
\ee
The conservation of probability
\be
\sum P_n(k)=1
\ee
then gives
\be
\vert A_0\vert^2=1-\omega_k.
\ee
Thus
\be
\vert R\vert^2=(1-\omega_k)\omega_k.
\ee
Using the expression of $\vert R\vert^2$ from equation (27) in (41) we find
\be
\omega_{\sc{k}}={ \exp{{-\pi\;(\,k_\bot^2 + m^2\,)} \over {qE_0}} \over 
{1+e^{{-\pi\;(\,k_\bot^2 + m^2\,)}\over {qE_0}} } },
\ee
\be
1-\omega_{\sc{k}}={1\over 
{1+e^{{-\pi\;(\,k_\bot^2 + m^2\,)}\over {qE_0}}}}.
\ee
The average number of pairs with wavenumber $\vec{k}$ is
\be
N(k)=\sum_{n=0}^{\infty}{n\, P_n(k)}={\omega_{\sc{k}}\over {1-\omega_{\sc{k}}}}
\ee
Using (42) and (43) in (44) we find
\beq
N(k)&=&
\exp{({{-\pi(k_\bot^2+m^2)}\over {qE_0}})} \nonumber \\
&\equiv & \vert\beta_k\vert^2\;\;(\rm{by \; definition}).
\eeq
This result of $\vert\beta_{\sc{k}}\vert^2$ is standard and give also the 
correct expression of vacuum persistence probability.
The probability that the vacuum remains the vacuum is given by
\beq
\vert <0, out\vert 0, in >\vert^2 &= & \prod_k \vert A_0 \vert^2, \nonumber \\
&=&\prod_k\;(\,1-\omega_{\sc{k}}\,),\nonumber \\
&=&\prod_k\;{1\over {1+\vert\beta_k\vert^2}},\quad (\rm{using\,(43)\,and \,(45)\,}),
\nonumber \\
&=&\exp{(-\sum_k{\ln{(1+\vert\beta_k\vert^2)} })}
\eeq
The equation (46) delivers the probability to find no pairs at future times.
From (44) and (45), we see that though our expression for $\vert R\vert$ 
differs from \cite{bar:ictp}, the results for Bogolubov co-efficient and the vacuum 
persistence probability turn out to be the same as the standard result of 
Schwinger \cite{sch:pr}. In our approach
\beq
\vert\beta_k\vert^2&=& e^{({{-\pi(k_\bot^2+m^2)}\over {qE_0}})} \nonumber \\ 
\vert\alpha_k\vert^2&=& 1+e^{({{-\pi(k_\bot^2+m^2)}\over {qE_0}})} .
\eeq
The results (44) - (47) suggest that the identification of Bogolubov co-efficients 
from the unitarity or charge conservation relation 
$\vert R\vert^2+\vert T\vert^2 =1$ is not a judicious choice. Replacing the 
summation in (46) by integration, we get exactly as in \cite{bar:ictp}
\be
\vert <0, out\vert 0, in >\vert^2 = 
\exp{({{-ELT}\over {2\pi}}\ln{(1+e^{{-m^2\pi}\over E})})},
\ee
in $(1+1)$ dimension and
\be
\vert <0, out\vert 0, in >\vert^2 = 
{{-E^2TV}\over {4\pi^2}}\ln{(1+e^{{-m^2\pi}\over E})},
\ee
in $(1+3)$ dimension. We understand that Padmanabhan \cite{pad:pri} has obtained
related results in this context and will be publishing them seperately.
\par
Now the essence of the calculation establishes that 
\par
\noindent
i) the complex paths taken in CWKB are all allowable and good paths and takes 
into account all the quantum mechanical contributions within the framework of 
semiclassical approximation and
\par
\noindent
ii)the interpretation of pair production in CWKB is a gauge independent result.
\par
\noindent
The equality of (46) and (45) with the Schwinger result raises a question. 
In \cite{bar:ictp} as well as in \cite{sri:iuc}, one also obtains the same results using (35), though
having the same answer for Bogolubov co-efficients and vacuum persistence 
probability. The difference occurs due to analytic continuation. When one goes
from 
$\psi_{\sc{T}}$ at $x\rightarrow +\infty$ to
$\psi_{\sc{I}}\,+\,R\psi_{\sc{R}}$ at $x\rightarrow -\infty$, not all the complex 
paths are taken into account in (14) and (15), whereas the CWKB does it correctly. 
We like to place the details in future; however similar view will also be found   
in \cite{nar:leb}.
\smallskip
\section{\bf{Particle Production In de-Sitter Spacetime}}
In standard black hole spacetime, of which de-Sitter is an example, particle 
production was discovered by Hawking \cite{har:prd}
using semi-classical analysis. In his method, the semi-classical propagator of
a scalar field propagating in Schwarzchild spacetime is analytically continued
in the time variable t to complex values. Using time dependent gauge in 
de-Sitter spacetime we obtained \cite{bis:cqg} that the appropriate vacuum for an inflationary 
early universe is the Hartle-Hawking vacuum and this vacuum acts as a blackbody 
of temperature $T_H ={ H/{2\pi}}$, the standard result.
It is therefore necessary to obtain the same result in space dependent gauge for a 
de-Sitter spacetime. 
\par
Consider a spacetime
\be
ds^2=B(r)dt^2-B^{-1}(r)dr^2-r^2(d\theta^2+\sin^2\theta)d\phi^2,
\ee
where 
$B(r)=1-H^2r^2$, 
in de-Sitter spacetime. For massless Klein-Gordon equation we have
\be
\Box\phi=0.
\ee
In the spacetime (50), with 
\be
\phi=\psi(r, t) Y^{m}_{l}(\theta,\phi),
\ee
the equation (51) reads
\be
{1\over {B(r)}}{{\partial^2\psi}\over {\partial t^2}} -
{1\over r^2}
{\partial\over {\partial r}}(r^2B(r){{\partial\psi}\over {\partial r}}) 
+{{l(l+1)}\over r^2}\psi =0.
\ee
Making the ansatz
$\psi=\exp{(-i\omega t)}f_{\sc{\omega}}$
in (53) we get
\be
{{B(r)}\over r^2}
{\partial\over {\partial r}}(r^2B(r){{\partial f_{\sc{\omega}}}\over {\partial r}}) 
+(\omega^2 -{{l(l+1)}\over r^2}B(r))f_{\sc{\omega}} =0.
\ee
Introducing 
$dr^*={{dr}\over {r^2B(r)}}$
in (54), we find
\be
{{\partial^2 f_{\sc{\omega}}}\over {\partial r^{*2}}}
+r^4(\omega^2 -{{l(l+1)}\over r^2}B(r))f_{\sc{\omega}} =0.
\ee
This is like our one dimensional Schroedinger equation in previous section with
\be
S(r)=\int^{r}
+r^2(\omega^2 -{{l(l+1)}\over r^2})^{1/2}{{dr}\over {r^2B(r)}}.
\ee
For $B(r)=1-H^2r^2$, the horizon is at $R=H^{-1}$ and the turning points are at 
$r=\pm\infty$. The $r\rightarrow -\infty$ has to be identified at the left of 
the horizon. The method of CWKB is now applied with the turning point at 
$r_1=+X,-X$, where $X$ is sufficiently large. As the particle enters the horizon 
radius, it moves in complex $(t, r)$-plane and we apply CWKB in the following way. 
we now write as before
\be
f_{\sc{\omega}}(x)\matrix{\rightarrow \cr x\rightarrow -\infty}
e^{-2iS(-x,-A)}+R\,e^{-2iS(-x,-A)},
\ee
where
\be
R=-{{ie^{-2iS(-X,-A)}}\over {1+e^{-2iS(X,-X)}}}.
\ee
Starting from $-A > - X$, as we proceed towards the right at the other 
turning point $+ X$ we have to cross the horizon.
Therefore reaching near horizon we move in upper half complex plane in a 
semi-circular path and back on the real axis. Thus
\be
S(X,-A)=\int^{-\varepsilon}_{-A}+\int_{S.C}+\int^{X}_{\varepsilon}.
\ee
Here A is an arbitrary point.

In (59), we take $l=0$ as it gives the most contribution 
(see ref.\cite{bar:ictp}) so that 
\be
S(x_1,X_2)=\int^{x_2}_{x_1}{{\omega}\over {B(r)}} dr.
\ee
As is evident from (58), the contribution from $A$, i.e.,
\be
S(A)=\int^{A}{{\omega}\over {B(r)}} dr
\ee
multiplies (58) as $ \exp{[iS(A)]}$ and hence we do not take its contribution in 
(59) for calculating $R$. We evaluate (59) and find 
\be
\int^{X}_{-A}\equiv-{{i\pi\omega}\over {2H}}+\delta_1,\;\;
\int^{X}_{-X}\equiv-{{i\pi\omega}\over {H}}+\delta_2,\;\;
\int_{S.C}=0,
\ee
where $\delta_1$ and $\delta_2$ are real. Hence
\be
\vert R\vert={{\exp{({{-\pi\omega}\over H})}}
\over {1+\exp{({{-2\pi\omega}\over H})}}}.
\ee
Thus we recover the Planckian spectrum with $T={H\over {2\pi}}$. This result is
remarkable. Thus we arrive at gauge invariant result even in CWKB.
\smallskip
\section{\bf{Discussion}}
Using CWKB we have obtained the gauge invariant result even in curved spacetime 
similar to electromagnetic example. It has been argued that \cite{sri:iuc}
the tunneling picture given in the present work fails in curved spacetime and 
in blackhole spacetime. However in the present work we judiciously circumvent 
the claim and obtained the Hawking result very nicely. We mention here the important 
results and would shortly explore the details.

It is worthwhile to mention some comparison with the standard works. Suppose
we neglect the repeated reflection between the turning points and determine the 
Bogolubov coefficients from (41). We find, neglecting the denominator in (27)
\be 
\vert R\vert^2=exp(-\pi M)
\ee
where
\[ M=\frac {\sqrt{k_\bot^2+m^2}}{qE_0}\]

We now calculate $\omega_k$ from (41). We find

\be 
\omega_k={1 \over 2} \left( 1 \pm \sqrt{1-4exp(-\pi M)}\right)
\ee
\be
1-\omega_k={1 \over 2} \left( 1 \mp \sqrt{1-4exp(-\pi M)}\right)
\ee

Taking the negative sign before the square root for $\omega_k$, we get
from (44) and (45) 
\be
\vert \beta_k\vert^2 \simeq {exp(-\pi M) \over 1- exp(-\pi M)}
\ee
provided $exp(-\pi M) \ll 1$.

Hence
\be 
\vert \alpha_k \vert^2= 1+\vert\beta_k\vert^2=\frac {1}{1-exp(-\pi M)},
\ee
so that  
\be
\vert \beta_k\vert^2 = \vert \alpha_k \vert^2 exp(-\pi M)
\ee
 
Carrying out this calculation for de-Sitter and Schwarzchild spacetime, we find
similar results. For de-Sitter spacetime, we get $T={H \over 2\pi}$ and 
for Schwarzchild we get $T={1 \over 8\pi M}$ in (1+3) dimensions. 

This exercise also exemplifies the usefulness of CWKB that takes more quantum
corrections than the other methods and gives a partial answer to the analytic
continuation, $x \rightarrow -\infty$ to $x \rightarrow +\infty$, ensuring that
the method of CWKB is more accurate and transparent, from the standpoint of physical
arguments, than the other methods.

\smallskip
\section{\bf{Acknowledgment}}                          
\par
A.Shaw acknowledges the financial support from ICSC World Laboratory,
LAUSSANE during the course of the work.
\end{document}